# Variationally Optimized Free Energy Flooding for Rate Calculation

James McCarty,[1] Omar Valsson,[1,2] Pratyush Tiwary,[3] and Michele Parrinello[1,2,*]

[1]*Department of Chemistry and Applied Biosciences, ETH Zurich and Facoltà di Informatica,*
*Instituto di Scienze Computationali, Università della Svizzera italiana,*
*Via Giuseppe Buffi 13, CH-6900, Lugano, Switzerland.*
[2]*National Center for Computational Design and Discovery of Novel Materials MARVEL.*
[3]*Department of Chemistry, Columbia University, New York, NY 10027.*
(Dated: September 17, 2018)

We propose a new method to obtain kinetic properties of infrequent events from molecular dynamics simulation. The procedure employs a recently introduced variational approach [Valsson and Parrinello, Phys. Rev. Lett. **113**, 090601 (2014)] to construct a bias potential as a function of several collective variables that is designed to flood only the associated free energy surface up to a predefined level. The resulting bias potential effectively accelerates transitions between metastable free energy minima while ensuring bias-free transition states, thus allowing accurate kinetic rates to be obtained. We test the method on a few illustrative systems for which we obtain an order of magnitude improvement in efficiency relative to previous approaches, and several orders of magnitude relative to unbiased molecular dynamics. We expect an even larger improvement in more complex systems. This and the ability of the variational approach to deal efficiently with a large number of collective variables will greatly enhance the scope of these calculations. This work is a vindication of the potential that the variational principle has if applied in innovative ways.

Molecular dynamics (MD) simulation can provide direct physical insight into the time evolution of molecular systems, and has become an important tool in many branches of science. However, the energy landscape of many interesting systems is characterized by numerous metastable states separated by large kinetic barriers [1]. A longstanding issue in the field of computational physics has been understanding phenomena which occur on such landscapes and involve timescales much longer than those accessible by traditional MD. This is arguably the most serious limitation of this powerful technique. While several methods have been developed to enhance the sampling of rare events [2–13] and obtain a static picture of the underlying landscape, calculating quantitative kinetic and dynamic properties remains a challenge.

Within the framework of transition-state theory (TST), one can envision adding a bias potential to enhance barrier crossing events out of local energy minima. The effect of the bias on the transition rates can then be rescaled by considering the theory of activated processes [14]. This is the idea behind many accelerated dynamics methods, such as conformational flooding [15, 16], hyperdynamics [17, 18], accelerated MD [19] and several other methods [20–28]. With a well-designed bias potential, one can obtain substantial speed-ups relative to unbiased MD; however, care must be taken to ensure that the bias vanishes at the transition state and that the system remains within local minima long enough to accumulate local averages. In other words, transitions in the biased ensemble should be representative of those that would be eventually sampled in the unbiased case in the long time limit.

Thus, the key problem in these approaches has been designing a bias potential that leaves the transition states untouched, remains simple to evaluate, yet gives acceleration relative to unbiased MD that scales well with the number of degrees of freedom. A poorly designed bias potential can either lead to vanishing boosts relative to MD or give inaccurate timescales [29, 30]. This is because the true high-dimensional potential energy surface has an enormously large number of low-lying saddle points, and the likelihood that a bias potential disturbs some of these saddle points in a non-trivial way becomes significant as the dimensionality increases [29, 30]. An appealing alternate approach is to shift the attention from dynamics on a potential energy surface to a free energy surface (FES), which is a coarse-grained representation of the system in terms of a few collective variables (CVs). Provided that the CVs distinguish between the various metastable states or basins, one then enhances fluctuations in these CVs with an appropriate bias to escape local minima.

Along these lines, Tiwary and Parrinello have recently proposed a method to recover the correct dynamics from biased metadynamics simulations [22, 31]. Using only a few relevant CVs they are able to recover correct transition rates by limiting the deposition frequency of the time-dependent bias. Their key idea is to bias the system perturbatively such that the frequency of perturbation is between the fast intra-basin and the slow inter-basin relaxation frequencies, thereby creating bias-free transition states. However, the need to reduce the bias deposition frequency is a severe computational handicap and forces one to rather lengthy calculations. In spite of this, the method has been used to calculate the unbinding kinetics of ligand/protein interactions and other applications, demonstrating the power of using dimensionality reduc-

tion [32–35].

In this Letter we propose a new method based on the efficient construction of a static bias which acts on a few relevant CVs and only acts below a threshold free energy, $F_c$, above which the FES is unaffected. The problem of estimating the free energy and constructing the bias is accomplished within the framework of a novel variational approach of Valsson and Parrinello [36]. Using this variational approach and including a constraint that ensures the bias potential will not act on the transition state, we obtain a static bias which accelerates transitions between metastable states, but makes no assumptions about the transition pathway or reaction coordinate. Importantly, no *a priori* assumption is made of the shape of the FES, which may be of arbitrary complexity. *Nota bene*: Flooding the FES, which is defined in terms of well-chosen CVs, is in no way the same as filling or lifting up all states below a certain potential energy as has been done in the past [19, 37].

In the following, we assume that there is a set of CVs, which map a given atomistic configuration, $\mathbf{R}$, onto a finite set of bounded coarse-grained variables, $\mathbf{s} \equiv \{s_i(\mathbf{R})\}$. The FES is then defined up to an irrelevant arbitrary constant as,

$$F(\mathbf{s}) = -\frac{1}{\beta} \log \int d\mathbf{R}\, \delta(\mathbf{s} - \mathbf{s}(\mathbf{R})) e^{-\beta U(\mathbf{R})}, \quad (1)$$

where $U(\mathbf{R})$ is the interaction potential and $\beta = 1/k_\mathrm{B}T$. We seek to introduce a bias potential, $V(\mathbf{s})$ that acts on the CVs such that the bias potential will enhance fluctuations out of local minima of the FES, while not biasing the transition state ensemble. The problem of finding the correct bias potential that will locally "flood" the FES is solved by minimizing the following functional of Valsson and Parrinello [36],

$$\Omega[V] = \frac{1}{\beta} \log \frac{\int d\mathbf{s}\, e^{-\beta[F(\mathbf{s})+V(\mathbf{s})]}}{\int d\mathbf{s}\, e^{-\beta F(\mathbf{s})}} + \int d\mathbf{s}\, p(\mathbf{s}) V(\mathbf{s}), \quad (2)$$

in which $p(\mathbf{s})$ is a chosen normalized probability distribution. The functional, $\Omega[V]$ is closely related to the relative entropy [38] and to the Kullback-Leibler divergence [39]. It can be shown that $\Omega[V]$ is a convex functional and that it has a global minimum that satisfies the following relation, valid up to a constant, [36],

$$V(\mathbf{s}) = -F(\mathbf{s}) - \frac{1}{\beta} \log p(\mathbf{s}), \quad (3)$$

when $p(\mathbf{s}) \neq 0$ and $V(\mathbf{s}) = \infty$ otherwise. At the minimum the CVs will be sampled according to the target distribution, $p(\mathbf{s})$, so we can tailor the sampling through the choice of $p(\mathbf{s})$. This is a great advantage of the variational formalism.

We now choose a variational form for $V(\mathbf{s})$ that together with an appropriate choice of $p(\mathbf{s})$ automatically ensures that the bias is zero when $F(\mathbf{s})$ is greater than a preassigned value, $F_c$, relative to the minimum of $F(s)$. To this effect we write $V(\mathbf{s})$ in the form

$$V(\mathbf{s}) = v(\mathbf{s}) \mathcal{S}(-v(\mathbf{s}) - F_c) \quad (4)$$

where $v(\mathbf{s})$ is the function that is allowed to vary in order to minimize $\Omega[V]$. $\mathcal{S}(x)$ is a sigmoidal switching function such that $\mathcal{S}(x) = 1$ for $x \to -\infty$ and $\mathcal{S}(x) = 0$ for $x \to \infty$. We also assume that $\mathcal{S}(x)$ is continuous and differentiable while still approximating a step function behavior. Within the local minimum, $v(\mathbf{s}) \approx -F(\mathbf{s})$. Such a functional form ensures that the bias is always less than $F_c$. A schematic depiction of the definition of the bias and switching function is presented in Fig. 1

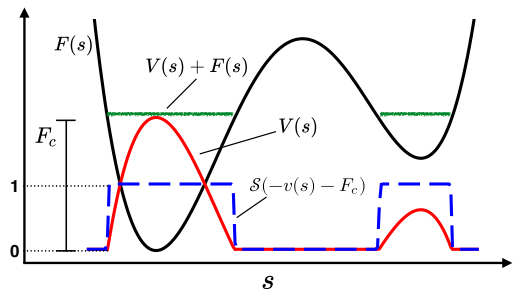

FIG. 1. A hypothetical one dimensional free energy surface, $F(s)$ is shown in black along with the switching function, $\mathcal{S}(-v(s) - F_c)$ (dashed blue line) for a chosen cutoff, $F_c$. The resulting bias, $V(s)$ is depicted by the red solid line, and fills $F(s)$ up to the value of $F_c$ (green solid line).

We note that the proof of the convexity of $\Omega[V]$ in Ref. [36] does not make any assumptions regarding the exact form of the bias potential, so this property of $\Omega[V]$ is fully preserved in the presence of the switching function. Of course, since we have limited the variational flexibility of $V(\mathbf{s})$, the reconstruction of the FES from Eq. 3 will be only approximate. However, this is of no relevance here since our only purpose is to construct a $V(\mathbf{s})$ that facilitates exit from the metastable state and does not affect the transition region.

We also would like to choose the target distribution such that for free energies below the cutoff, $F_c$, the bias compensates almost exactly the underlying free energy and is zero otherwise. This can be achieved if we have a good estimate of the free energy surface $F^*(\mathbf{s}) \cong F(\mathbf{s})$ and write $p(\mathbf{s})$ as:

$$p(\mathbf{s}) = \frac{\mathcal{S}(F^*(\mathbf{s}) - F_c)}{\int d\mathbf{s}'\, \mathcal{S}(F^*(\mathbf{s}') - F_c)}. \quad (5)$$

This choice selects only those portions of $s$-space in which $F^*(\mathbf{s}) \leq F_c$. Of course $F^*(\mathbf{s})$ is not *a priori* known, but can be estimated in a self-consistent procedure in which we first guess $F^*(\mathbf{s})$, then minimize $\Omega[V]$ using the corresponding $p(\mathbf{s})$ to obtain a new $V(\mathbf{s})$, which in turn gives a new estimate of $F^*(\mathbf{s})$ from $F^*(\mathbf{s}) \cong -v(\mathbf{s})$. The process is iterated until a converged result is reached, in a way very similar to that described in Ref. [40] in a related but different context. One way of making this variational principle practical is to expand $v(\mathbf{s})$ in a basis function set, $f_k(\mathbf{s})$,

$$v(\mathbf{s}) = \sum_k \alpha_k \cdot f_k(\mathbf{s}), \qquad (6)$$

and use the expansion coefficients, $\{\alpha_k\}$, as variational parameters [36, 40]. The set of basis functions may be plane waves, Chebyshev polynomials, or any other suitable basis set, depending on the nature of the CVs and the problem at hand. Of course any other variational form for $v(\mathbf{s})$ can equally well be chosen. For the optimization of $\Omega[V]$ with respect to the set of variational parameters $\{\alpha_k\}$ we employ the stochastic optimization method introduced in Ref. [41], as done previously in Refs. [36, 40]. In this work, we have employed a Fermi-type switching function for $\mathcal{S}(x)$,

$$\mathcal{S}(-v(\mathbf{s}) - F_c) = \frac{1}{1 + e^{\lambda[-v(\mathbf{s}) - F_c]}}, \qquad (7)$$

where $\lambda$ is a parameter with units of inverse energy that determines how quickly the switching function goes to zero. The stochastic optimization procedure is presented in more detail in the SM.

After the minimization procedure, the bias, $V(\mathbf{s})$ is implemented as a fixed bias which speeds up the molecular dynamics time by facilitating escape over free energy barriers. The physical time is recovered using the relationship [17, 19],

$$t^* = \sum_i^{n_{tot}} \Delta t_i^* = \Delta t_{\mathrm{MD}} \sum_i^{n_{tot}} e^{\beta V(\mathbf{s})}$$
$$t^* = t_{\mathrm{MD}} \left\langle e^{\beta V(\mathbf{s})} \right\rangle_V, \qquad (8)$$

where $\Delta t_{\mathrm{MD}}$ is the MD integration time step, $t_{\mathrm{MD}}$ is the time as measured in a biased molecular dynamics run, and $t^*$ is the "real" time in an unbiased simulation corresponding to $t_{\mathrm{MD}}$. The quantity $\left\langle e^{\beta V(\mathbf{s})} \right\rangle_V = \frac{t^*}{t_{\mathrm{MD}}}$ is the acceleration factor which is a measure of how much the time is boosted, and the subscript $V$ denotes sampling under the biased potential.

We now proceed with a few illustrative examples. First, we consider the classical case of the $C_{7eq} \rightarrow C_{7ax}$ conformational change of alanine dipeptide in vacuum, which can be distinguished by the two backbone dihedral angles, $(\phi, \psi)$. The apparent barrier height for the forward conversion is approximately 34 kJ/mol. We bias both angles $\phi$ and $\psi$, and use a plane wave expansion, $v(\mathbf{s}) = \sum_k \alpha_k e^{i\mathbf{k}\cdot\mathbf{s}}$. The procedure has been implemented in a private development version of the PLUMED 2 plug-in [42], which will be made available in the near future.

Details of the computational methods and optimization procedure along with the resulting bias potentials are presented in the SM. To test the efficiency of the method, we perform a short optimization step of the bias for several different $F_c$ values, which gives us a series of bias potentials of increasing strength. Subsequently, we used each potential as a fixed bias for 60 independent trajectories all started from the same initial $C_{7eq}$ configuration, from which we measured the first passage time (fpt) into the $C_{7ax}$ state. As a comparison we also perform well-tempered metadynamics (WTMetaD) with similar parameters to those used in Ref. [22] (see SM).

Fig. 2(a) shows the cumulative distribution of the unrescaled first passage times for increasing $F_c$ values as compared to WTMetaD at $T = 250$ K. For $F_c$ greater than 24 kJ/mol, the free energy flooding method drives barrier crossing faster than WTMetaD. This is due to the inferior efficiency of WTMetaD to fill the basin relative to the variational approach. This lack of efficiency in WTMetaD is made worse by the need of depositing the Gaussians at a low pace. The optimization of the bias using the variational approach only needs to be performed once for each $F_c$ and converges rapidly (within 1 ns) to an effective bias. As shown in the SM, the MD time needed for the optimization step is negligible as compared to the total MD time needed for the subsequent production runs and thus can be ignored when considering the efficiency of our approach.

Fig. 2(b) shows the fpt distributions after rescaling the time according to Eq. 8. All of the transition times collapse onto a single distribution, which corresponds to the unbiased trajectory distribution. This indicates that even for values of $F_c$ which are just a few $k_\mathrm{B}T$ below the apparent free energy barrier, the method recovers the correct distribution of passage times. Fig. 2(c) shows the mean first passage times as a function of $F_c$ along with the associated acceleration factor. While the acceleration factor increases by several orders of magnitude with increasing $F_c$, due to the exponential dependence of the rate on the barrier height, the rescaled mean first passage times are constant within the error bars and are in quantitative agreement with the first passage time from unbiased MD of $23 \pm 3.8$ $\mu s$. We expect that a test similar to the one illustrated in Fig. 2(c) will be extremely useful in ascertaining the reliability of the accelerated timescales in complex systems. Sensitivity of the rescaled escape rates to the extent of flooding is typically a clear-cut indication that the chosen collective variable is not equilibrating fast enough, or that the transition states have been corrupted [30].

As a second example we construct a model of benzophenone in water. Since the true system is adequately sampled by unbiased molecular dynamics we have stiffened the torsions around the ketone group. This stiff model allows us to treat the back and forth out-of-plane bending motion of the two phenyl rings around the ketone in the presence of molecular solvent, shown in Fig. 3(a), as a rare event. In Fig. 3(b), the FES of this motion obtained using the variational method with dihedral angles $\phi_1$ and $\phi_2$ is shown for the transition between the two conformations. The FES shown in Figure 3(b) is only a portion of the total FES in the space of the dihedral angles. Since we are only interested in a limited region of the FES, we use products of Chebyshev polynomials defined in the range $[-1 : 1]$ as basis functions, which allows us to construct a bias confined to the region of conformational space for which we are interested. Computational details are presented in the SM. The apparent barrier in Fig. 3(b) is 22 kJ/mol. Using a fixed value of $F_c = 12$ kJ/mol we obtain a bias which partially floods both wells and gives a boost in the kinetics, corresponding to a new barrier of $\sim 4\, k_{\rm B}T$. In order to ensure a stable optimization, we use 4 multiple walkers during the optimization step. The rates of the forward and reverse transition after rescaling the trajectory using Eq. 8 are compared in Fig. 3(c) to those from unbiased simulations performed at different temperatures. Within the error bars, the forward and reverse rates are equal as expected.

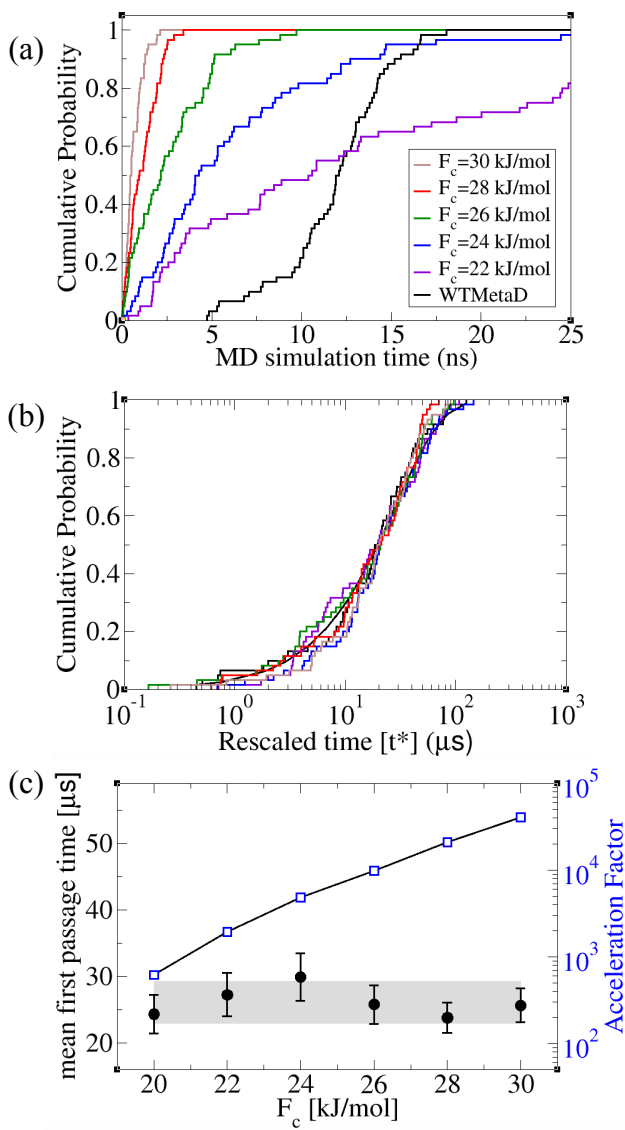

FIG. 2. (a) Cumulative distribution of the unrescaled first passage times for the $C_{7eq} \rightarrow C_{7ax}$ transition in alanine dipeptide in vacuum at $T = 250$ K using different $F_c$ values (From left to right: $F_c = 30, 28, 26, 24,$ and $22$ kJ/mol) as compared to WTMetaD (black). (b) Cumulative distribution of first passage times after time rescaling. As a guide to the eye, an exponential fit to the data with $\tau = 28.2$ $\mu s$ is shown (black line). (c) Mean first passage times plotted vs. increasing $F_c$ (left axis). The shaded gray region depicts the standard error determined from WTMetaD. Also shown is the corresponding acceleration factor (blue) on a logarithmic scale (right axis).

In conclusion, in this Letter we have presented an efficient method to obtain kinetic properties of rare events using a variational procedure to construct a static bias in free energy space. The method is general and easy to implement, making it suitable for a variety of applications. We have demonstrated the procedure on two model systems and have shown that at least one order of magnitude of efficiency can be gained relative to WTMetaD [22], and several orders of magnitude relative to unbiased MD. More impressive gains are to be expected in more complex systems. With this new method, many other computational advantages are possible. Since we are able to focus separately on one particular metastable state, we can limit our attention to this particular state and use collective variables appropriate for this state. This is a far less demanding requirement than finding collective variables able to reconstruct the whole configuration space. Furthermore, we can increase the number of collective variables since the space to be filled is small. Approximate but highly dimensional forms of bias can also be used. This considerably alleviates the problem, sometimes vexing, of finding appropriate CVs. All of these considerations make the method preferable not only to WTMetaD, but to a more straightforward approach in which the global free energy is calculated first and only later the states below $F_c$ are selected. This work is also a clear example of how the variational property of $\Omega[V]$ can be employed in a novel manner.

J.M., O.V., and M.P. acknowledge funding from the National Center for Computational Design and Discovery of Novel Materials MARVEL and the European Union Grant No. ERC-2009-AdG-247075. We would like to thank Matteo Salvalaglio for providing the benzophenone model parameters. All calculations were performed on the Brutus HPC cluster at ETH Zurich.

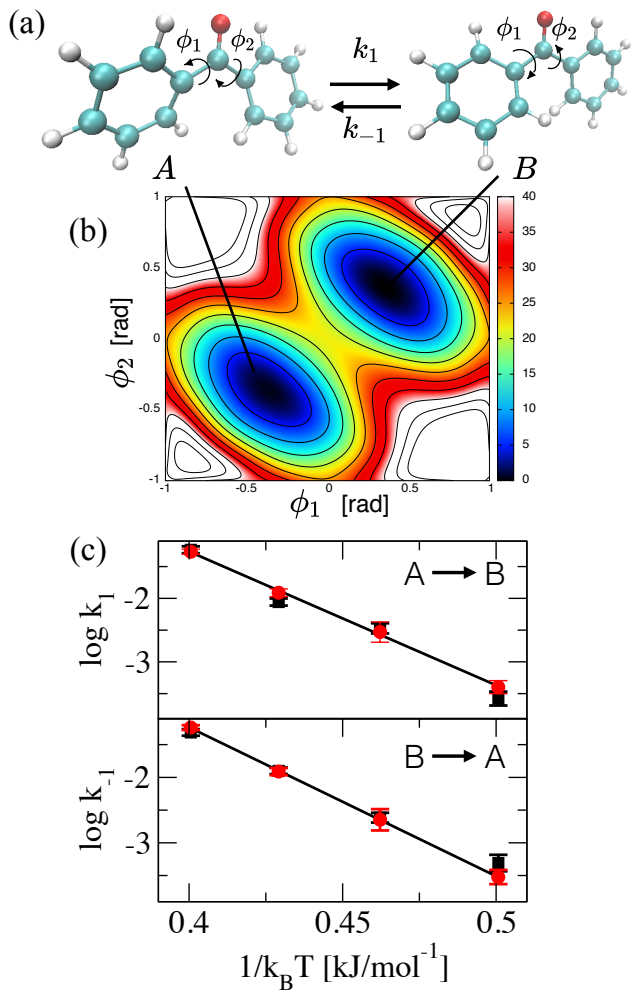

FIG. 3. (a) Representative conformers and the definition of the CVs for the stiffened benzophenone model. (b) Reference free energy surface (kJ/mol) for the region of configuration space of interest. (c) Arrhenius plot of the rate of the interconversion from conformer $A \to B$ (upper) and $B \to A$ (lower). Rates obtained from biased trajectories using $F_c = 12$ kJ/mol are show with red circles and unbiased MD are shown in black squares. The line is a linear fit to the biased (red circle) data. The slope gives an estimate of $21.0 \pm 0.5$ kJ/mol (upper) and $22.8 \pm 0.14$ kJ/mol (lower) for the barrier height which are both in agreement with the apparent barrier of $\sim 22$ kJ/mol obtained from enhanced sampling.